\documentclass[aps,pra,twocolumn,showpacs,amsmath,amssymb,floatfix,superscriptaddress]{revtex4-1}
\usepackage[usenames,dvipsnames,svgnames,table]{xcolor}
\usepackage{graphicx}
\usepackage[bookmarks,bookmarksopen,bookmarksnumbered,colorlinks,linkcolor=red,linktocpage,citecolor=blue,urlcolor=cyan,pdfpagemode=UseOutline]{hyperref}
\usepackage{epstopdf}

\def\bea{\begin{eqnarray}}
\def\eea{\end{eqnarray}}
\def\ben{\begin{equation}}
\def\een{\end{equation}}
\def\benu{\begin{enumerate}}
\def\enu{\end{enumerate}}

\def\bei{\begin{itemize}}
\def\eei{\end{itemize}}
\def\benu{\begin{enumerate}}
\def\enu{\end{enumerate}}

\def\sss{\scriptscriptstyle\rm}

\def\1var{(\bx_1...\bx\N)}

\def\br{{\bf r}}

\def\bx{{x}}

\def\c{_{\sss C}}

\def\xc{_{\sss XC}}
\def\Hx{_{\sss HX}}
\def\Hxc{_{\sss HXC}}

\def\N{_{\sss N}}
\def\H{_{\sss H}}

\def\sph_int{ {\int d^3 r}}

\newcommand{\intd}{\mathrm{d}}
\newcommand{\vect}[1]{\mathbf{#1}}
\newcommand{\matelem}[3]{\left\langle #1 \left| #2 \right| #3 \right\rangle}

\providecommand{\abs}[1]{\left|#1\right|}

\newcommand{\parref}[1]{(\ref{#1})}

\def\wt{\texttt{w}}
\def\wtb{\overline{\wt}}
\def\Hxcw{_{{\sss HXC},\wt}}

\def\xcw{_{{\sss XC},\wt}}

\begin{document}

\title{Exact and approximate Kohn-Sham potentials in ensemble density-functional theory}
\author{Zeng-hui Yang}
\affiliation{Department of Physics and Astronomy, University of Missouri, Columbia, MO 65211, USA}
\author{John R. Trail}
\affiliation{Theory of Condensed Matter Group, Cavendish Laboratory, University of Cambridge, Cambridge CB3 0HE, United Kingdom}
\author{Aurora Pribram-Jones}
\affiliation{Department of Chemistry, University of California-Irvine, Irvine, CA 92697, USA}
\author{Kieron Burke}
\affiliation{Department of Chemistry, University of California-Irvine, Irvine, CA 92697, USA}
\author{Richard J. Needs}
\affiliation{Theory of Condensed Matter Group, Cavendish Laboratory, University of Cambridge, Cambridge CB3 0HE, United Kingdom}
\author{Carsten A. Ullrich}
\affiliation{Department of Physics and Astronomy, University of Missouri, Columbia, MO 65211, USA}
\date{\today}
\pacs{31.15.E-, 31.15.ee, 31.10.+z, 71.15.Qe}

\begin{abstract}
We construct exact Kohn-Sham potentials for the ensemble density-functional theory (EDFT) from the ground and excited states of helium. The exchange-correlation (XC) potential is compared with the quasi-local-density approximation and both single determinant and symmetry eigenstate ghost-corrected exact exchange approximations. Symmetry eigenstate Hartree-exchange recovers distinctive features of the exact XC potential and is used to calculate the correlation potential. Unlike the exact case, excitation energies calculated from these approximations depend on ensemble weight, and it is shown that only the symmetry eigenstate method produces an ensemble derivative discontinuity. Differences in asymptotic and near-ground-state behavior of exact and approximate XC potentials are discussed in the context of producing accurate optical gaps.
\end{abstract}

\maketitle

\section{Introduction}
The balance of useful accuracy with computational efficiency makes density-functional theory (DFT) popular for finding ground-state electronic properties of a wide range of systems and materials \cite{B12}. While exact conditions \cite{PBE96} and fitting to chemical data sets \cite{B93} are often used to construct approximations, another major source of inspiration has been highly accurate calculations of Kohn-Sham (KS) quantities for simple systems, such as the He atom \cite{UG94}.  The exact KS potential, orbitals, energies, and energy components have been enormously useful in illustrating basic theorems of DFT and testing approximations. Many algorithms now exist for extracting the KS potential from accurate densities \cite{LB94,GLB95,PVW03}.

Time-dependent density-functional theory (TDDFT) \cite{MMNG12,U12} has become the standard DFT method for calculating excitation energies, at least for molecules, with typical accuracies and efficiency comparable to what can be achieved in ground-state DFT \cite{JPCA10}. Once again, accurate KS energies, of both occupied and unoccupied orbitals, play a vital role \cite{AGB03}. But alternative density-functional approaches for excitation energies can be valuable, both as practical tools and for gaining physical insight
\cite{Gb99,LN99}. The ensemble density-functional theory (EDFT) formalism for excited states \cite{T79,HT85,T87,GOKb88,GOK88,OGKb88} is based on a variational principle of ensembles comprising the ground state and a chosen number of excited states. Despite its rigorous formal framework and appealing physical motivation\cite{N95,N98,N01,GPG02,PGP13,PP14}, the EDFT excited-state formalism has seen only limited practical success. The lack of good approximate exchange-correlation (XC) functionals for EDFT leads to inaccurate transition frequencies. Better approximations are needed for EDFT to become more useful.

Here, we 
describe an algorithm that extracts the ensemble KS and
XC potentials from the various eigenstate densities, and
apply that algorithm to highly accurate densities of the helium atom.
We use the exact results to analyze errors in approximations
that have been designed for use in EDFT, plot various potentials, and check the virial theorem
for the ensemble correlation potential.
We demonstrate the weight-independence of
transition frequencies in the exact case, but also find a strong
weight-dependence in the individual elements contributing to the
exact expression, all of which cancels in the final excitation energy.
We show that approximations all yield (incorrectly) weight-dependent
transition frequencies, and demonstrate how this is related to the ensemble derivative discontinuity.

\section{Theory}
An ensemble in EDFT consists of the ground state and $M$ excited states. For the lowest $M+1$ eigenstates $\Psi_m$ of the many-body 
Hamiltonian $\hat{H}$, sorted by energy in ascending order,
each state is assigned a weight $\wt_M$. EDFT states that for
\ben \wt_0 \ge \wt_1\ge \wt_2\ge\cdots\ge \wt_M\ge0, \een there is a one-to-one
correspondence between the ensemble density \ben
n(\br)=\sum_{m=0}^M \wt_m\matelem{\Psi_m}{\hat{n}(\br)}{\Psi_m}
\een and the external potential \cite{GOKb88,GOK88}. 
A Kohn-Sham (KS) scheme can then be constructed in the usual way
\cite{GOK88}. 

We consider only bi-ensembles of the ground and first-excited
states.  For a non-degenerate ground state,
\begin{align}
n_{\wt}(\br)&=\wtb~n_0(\br)+g~\wt~n_1(\br),~~~~~~\wt\leq1/(1+g)\\
E_{\wt}[n_{\wt}]&=\wtb~E_0+g~\wt~E_1,
\end{align}
where $g$ is the degeneracy of the excited
state, $\wtb=1-g\,\wt$, and subscripts 0 and 1 refer
to the ground and excited states.  
EDFT also holds for ensembles of states
that share a symmetry-projected Hamiltonian \cite{JG89}. 
For helium, the ground state is a singlet, the first excited
state is a triplet, and the second excited state is
again a singlet, shown in Fig. \ref{fig:He:density:compare}. 
The (unprojected) bi-ensemble always includes the ground state
and the first excited state.
Here we focus on calculations
in the spin-projected ensemble to find the transition
to the lowest singlet.

The corresponding ensemble KS potential
$v_{s,\wt}[n_{\wt}](\br)$ is defined as the potential of the
non-interacting system 
\ben
\left\{-\frac{1}{2}\nabla^2+v_s(\br)\right\}\phi_j(\br)=\epsilon_j\phi_j(\br),
\een
which reproduces the exact ensemble density as
\ben
n_\wt(\br)
=(1+\wtb)\abs{\phi_1(\br)}^2+g\, \wt\, \abs{\phi_2(\br)}^2,
\een
where $\phi_j(\br)$ are KS orbitals. Atomic units ($e=\hbar=m_e=1/4\pi\epsilon_0=1$) are
used throughout, and all KS quantities are $\wt$-dependent. Then
\ben
E_\wt[n]=T_{s,\wt}[n]+\int\intd^3r\;n(\br)v(\br)
+E\Hxcw[n],
\een
where
$T_{s,\wt}[n]=(1+\wtb)\, t_1 +g\, \wt\, t_2$
is the ensemble KS kinetic energy, with $t_j$ the kinetic
energy of $\phi_j$.  $v(\br)$ is the external potential of the interacting
system.
\bea
\label{eqn:EHx}
E\Hx&=\wtb~\matelem{\Phi_{0,\wt}[n]}{\abs{\br-\br'}^{-1}}{\Phi_{0,\wt}[n]}\notag\\
&+g~\wt\matelem{\Phi_{1,\wt}[n]}{\abs{\br-\br'}^{-1}}{\Phi_{1,\wt}[n]}
\eea
is the ensemble Hartree-exchange energy, and the ensemble correlation energy $E\c=E\Hxc-E\Hx$. $\Phi_{i,\wt}[n]$ is the KS many-body wavefunction, with $i=0$ or $1$ again indicating the ground or excited state. Here we choose $E\H$ to be the Hartree energy of the ensemble density, although
it contains ``ghost'' interactions\cite{GPG02}.  The exchange energy
is then defined as the expectation of the electron-electron
repulsion on the KS ensemble minus the Hartree energy.
This definition of $E\Hxc$ is consistent with our choice of 
spin eigenstates that are necessarily multi-determinant.  
The ensemble KS potential is 
\ben
v_{s,\wt}[n](\br)=v(\br)+v\Hxcw[n](\br),
\een 
where $v\Hxcw[n](\br)=\delta E\Hxcw[n]/\delta
n(\br)$. The excitation energy is then independent of $\wt$:
\ben
\omega=E_1-E_0=\Delta \epsilon_\wt+{\partial
E\Hxcw[n]}/{\partial \wt}|_{n=n_\wt},
\label{eqn:theory:exciteng} 
\een
where $\Delta \epsilon_\wt=\epsilon_{2,\wt}-\epsilon_{1,\wt}$.

The $\wt$-dependence of the HXC energy comes from both the $\wt$-dependence of $n_\wt(\br)$ and from the HXC energy functional. Eq. \parref{eqn:theory:exciteng} shows that the correction to the KS gap originates from the $\wt$-dependence of XC, not from $n_\wt(\br)$. Using a ground-state XC functional in EDFT yields no correction to the KS excitation energy. EDFT is a more general theory encompassing ground-state DFT, and the ground-state XC functional is only a special case ($\wt=0$) of the ensemble XC functional. However, the excitation energies can also be obtained from the difference of two consecutive equiensemble energies. In contrast to Eq. \parref{eqn:theory:exciteng}, the density-based $\wt$-dependence of $E\Hxcw$ does not drop out in that approach, and using ground-state XC functionals would yield finite corrections. These two approaches for the excitation energy yield the same result using the \emph{exact} functional, but no known approximations can achieve such consistency.

\section{Inversion Method}
The only unknown in the ensemble KS procedure is the XC functional. Without this functional, an inversion method for EDFT is needed to extract XC potentials from accurate densities. Ref. \cite{N95} presented an inversion scheme for EDFT similar to the van Leeuwen-Baerends (LB) algorithm in ground-state DFT\cite{LB94}, but we found its numerical stability unsatisfactory. Ref. \cite{PVW03} observed that a LB-type algorithm cannot change the local sign of the KS potential during the iteration. While not a fundamental problem, it makes the algorithm less stable. Also, it can be hard to obtain the $-1/r$ asymptotic behavior of $v\xc$ using the LB algorithm without having to build it in the initial guess. Ref. \cite{PVW03} suggested an alternative ground-state density-inversion algorithm, where the xc potential is updated iteratively by

\begin{align}
v\xc^{(i+1)}(r)&=v\xc^{(i)}(r)+\alpha r^\beta[n_{\text{KS}}^{(i)}(r)-n(r)]\notag\\
&\quad+[I_{\text{KS}}^{(i)}-I]\left[\theta(1-r)r^\gamma+\frac{\theta(r-1)}{r^\delta}\right],
\label{eqn:numerical:gsdeninv}
\end{align}
where $\alpha$, $\beta$, $\gamma$, $\delta$ are parameters controlling the speed of convergence, and $I$ is the ionization energy. In the asymptotic region, the density difference in the second term of Eq. \parref{eqn:numerical:gsdeninv} is very small, so the convergence needs to be accelerated by the use of the $r^\beta$ in front of this term. Even so, the $-1/r$ asymptotic behavior of $v\xc$ can be hard to obtain, and the third term of Eq. \parref{eqn:numerical:gsdeninv} is there to ensure this asymptotic behavior.

Our scheme for EDFT is based on the ground-state density-inversion method of Ref. \cite{PVW03} and Eq. \parref{eqn:numerical:gsdeninv}, producing the ensemble XC potential from any given ensemble density. For simplicity, we describe the scheme for spherical systems, but it can be extended to other systems easily. We modify the ground-state Eq. \parref{eqn:numerical:gsdeninv} for EDFT usage as
\ben
v_{\text{xc},\wt}^{(i+1)}(r)=v_{\text{xc},\wt}^{(i)}(r)+\alpha r^{\beta}[n_{\text{KS},\wt}^{(i)}(r)-n_{\wt}(r)]/h(r),
\label{eqn:numerical:ensdeninv}
\een
where $h(r)$ is described below. Since the ionization energies of Eq. \parref{eqn:numerical:gsdeninv} are not defined for an ensemble, a double-loop scheme is used to ensure the correct $-1/r$ asymptotic behavior.

In the first iterative loop, we update the ensemble xc potential with Eq. \parref{eqn:numerical:ensdeninv} and set $h(r)=1$. Convergence is reached when
\ben
\int\intd^3r\;\abs{n^{(i)}_{KS,\wt}(\vect{r})-n_\wt} <\Delta_1,
\label{eqn:numerical:convg}
\een
for a chosen accuracy $\Delta_1$. Even if large $\beta$ values are used to accelerate convergence in the large-$r$ region, this first loop is usually insufficient to produce the $-1/r$ asymptotic behavior in the ensemble xc potential, due to the exponential asymptotic decay of the density. To compensate for this, we use a second iterative loop. Starting from the result of the first loop, the ensemble xc potential is updated using Eq. \parref{eqn:numerical:ensdeninv}  with $h(r)=n_\wt(r)$ and new values of $\alpha$
 and $\beta$. The convergence of the second loop is also checked with Eq. \parref{eqn:numerical:convg}, but with a smaller $\Delta_2$. This second loop updates the ensemble xc potential with the relative error in the ensemble density, so the correction in the large-$r$ region for each iteration is larger than in the first loop. The second loop is therefore more sensitive to the initial guess than the first loop, so it cannot be used independently. We consistently obtain $-1/r$ asymptotic behavior in the ensemble XC potentials produced by this double-loop procedure, without having to build it in the algorithm or in the initial guess. This double loop scheme guarantees both numerical stability and good convergence in the asymptotic
 region.

For ensembles of the helium atom, we found that parameters $\alpha\in[0,2]$ and $\beta\in[0,2]$ guarantee convergence of the first loop. For the second loop, $\alpha\in[0,0.0001]$ and $\beta\in[0,2]$ guarantee convergence, if $\wt$ is not close to 0. As $\wt$ approaches 0, the value of $\alpha$ needs to be smaller to prevent the second loop from becoming unstable. The double-loop scheme has had good numerical performance in all types of grids and discretizations of the Hamiltonian tested thus far.

\section{Exact results for He atom}
\label{sec:exact:results}
\begin{figure}[htbp]
\includegraphics[width=0.9\columnwidth]{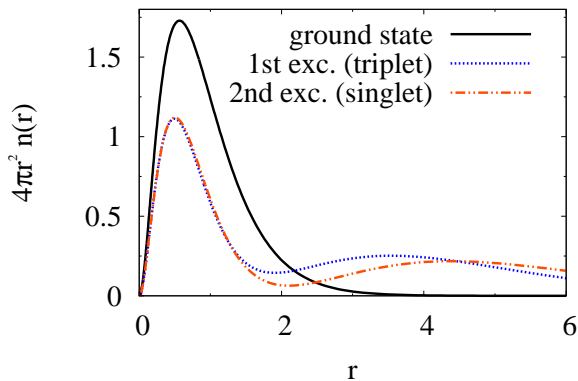}
\caption{Radial densities for the three lowest eigenstates of helium. Color online.}
\label{fig:He:density:compare}
\end{figure}
We apply this scheme to highly accurate helium densities. Fig. \ref{fig:He:density:compare} shows
the ground and first two excited state densities for helium, which are essentially numerically exact.
Two-body electronic wave functions were obtained by optimizing an
expansion in Hylleraas functions\cite{drake_94}.
Analytic integration of the density matrix
associated with the optimum wave function provides an accurate
spherically averaged charge density at each radius as a sum of terms.
Basis sets composed of $376$ and $406$ Hylleraas functions
for the singlet and triplet states, respectively, result in total
energies within $10^{-11}$ a.u. of accurate estimates\cite{nistor_04}. 
The errors in the virial are below $10^{-12}$ a.u. for the ground state
and $10^{-8}$ a.u. for the 
first
singlet excited state, used in the singlet bi-ensemble.
Our calculation for $\wt=0$ agrees
with the known exact ground-state DFT quantities of helium \cite{UG94}.

\begin{figure}[htbp]
\includegraphics[width=0.9\columnwidth]{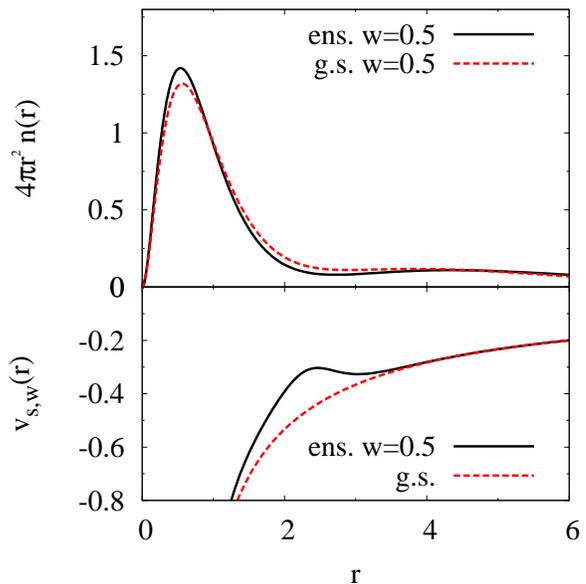}
\caption{Radial densities and KS potentials for helium in singlet EDFT. The black solid lines are equiensemble properties. The red dashed line in the upper panel shows an equiensemble density constructed from orbitals of the ground-state KS potential; the red dashed line in the lower panel shows the exact ground-state KS potential.}
\label{fig:He:general}
\end{figure}

\begin{figure}[htbp]
\includegraphics[width=0.9\columnwidth]{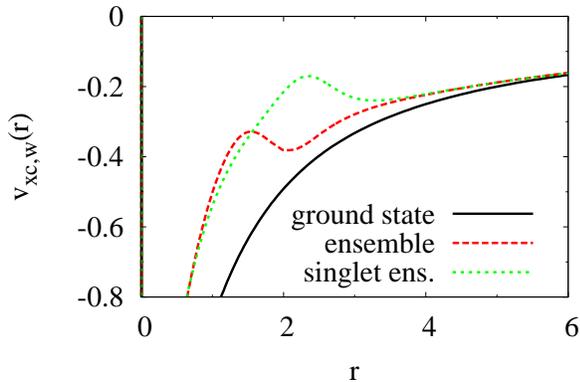}
\caption{XC potentials for the helium ground state, bi-ensemble, and symmetry-projected singlet ensemble, produced by inverting ensemble densities constructed from the states shown in Fig. \ref{fig:He:density:compare}.}
\label{fig:He:vxc:compare}
\end{figure}

The exact equiensemble density and potential are
plotted in Fig. \ref{fig:He:general}, along with those resulting 
from an equal mixture of orbitals from the ground-state KS potential. 
The subtle shell-like structure in
the ensemble density corresponds to the cross-over between the
ground-state density and the first singlet excited-state density.
The upward bump near $r=2.5$ in the ensemble KS potential
ensures its ensemble density matches the interacting one, unlike
the ensemble of orbitals from the ground-state KS potential. This bump is shifted left in the XC potential for the unprojected bi-ensemble (Fig. \ref{fig:He:vxc:compare}).

\begin{figure}[htbp]
\includegraphics[width=0.9\columnwidth]{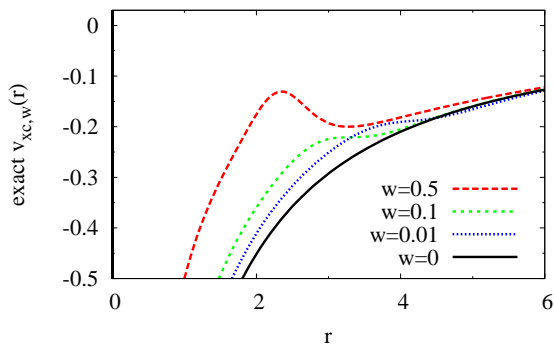}
\caption{The exact XC potential for the helium singlet ensemble at various ensemble weights.}
\label{fig:He:vxcatw}
\end{figure}

Fig. \ref{fig:He:vxcatw} shows the exact
ensemble XC potentials at various $\wt$ values, which have
been found by subtracting the Hartree potential
of the ensemble density from the KS potential. 
The bump near $r=2.5$ develops as $\wt$ increases. Even when $\wt$ is close to 0,  $v\xcw(r)$ differs from the $\wt=0$ (ground-state) XC potential in Fig. \ref{fig:He:vxcatw}. The potentials shift further and further from the ground-state curve in the small-$r$ region as $\wt$ increases.

\begin{figure}[htbp]
\includegraphics[width=0.9\columnwidth]{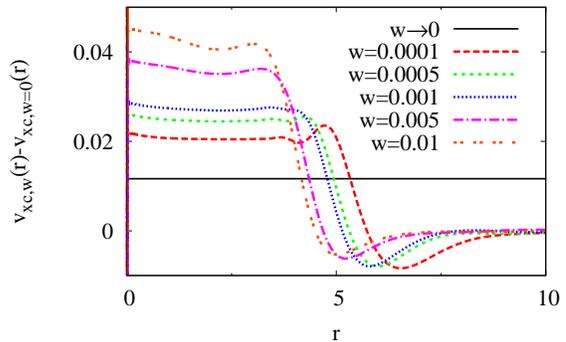}
\caption{The exact potential jump $\Delta v\xc$ as $\wt\rightarrow 0$. The location of the step depends logarithmically on $\wt$. As $\wt\rightarrow 0$, the drop-off to the $\wt=0$ value moves infinitely far from the origin.}
\label{fig:He:deltavxc:exact}
\end{figure}

\begin{figure}[htbp]
\includegraphics[width=0.9\columnwidth]{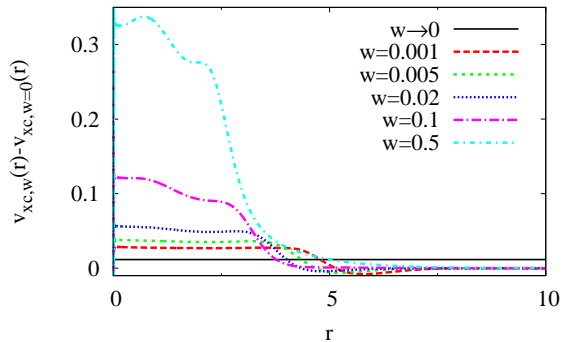}
\caption{The exact potential jump $\Delta v\xc$, showing the shoulder in the XC potential developing from the small-$\wt$ step as $\wt$ increases.  Since $\wt$ is no longer near zero, the asymptotic formula for the position of the drop-off no longer holds.}
\label{fig:He:deltavxc:more}
\end{figure}

This discrepancy between small-$\wt$ and $\wt=0$ potentials is due to the ensemble derivative discontinuity\cite{L95}. For any nonzero $\wt$, the asymptotic behavior of the ensemble density is dominated by that of the excited state. Levy \cite{L95} proved an analog of the derivative discontinuity of ground-state DFT: the ensemble KS highest-occupied-molecular-orbital (HOMO) energy has a finite change as $\wt$ changes from 0 (ground state) to $0_+$:
\begin{align}
\label{eqn:numerical:deltavxc1}
\Delta v\xc(\br)&=\lim_{\wt\to0}v\Hxcw[n_\wt](\br)-v\Hxc[n](\br)\\
&=\lim_{\wt\to0}{\partial E\Hxcw[n]}/{\partial \wt}|_{n=n_\wt}.
\label{eqn:numerical:deltavxc2}
\end{align}
This is an exact property of number-conserving excitations\cite{note1}. According to Eq. \parref{eqn:theory:exciteng} and \parref{eqn:numerical:deltavxc2}, we obtain $\Delta v\xc=0.0116$ a.u. for the singlet bi-ensemble.

Fig. \ref{fig:He:deltavxc:exact} shows the exact XC potential jump for small $\wt$ values. A step structure occurs since the ensemble density at small $r$ is dominated by the HOMO density, and at large $r$ the dominating behavior switches to the lowest-unoccupied-molecular-orbital (LUMO) density, which decays more slowly than the HOMO density. As $\wt$ decreases, the switching point $r\c$ moves to the right. In the limit of $\wt\to0$, the HOMO density dominates $n_\wt(r)$ for finite $r$, so $\Delta v\xc(r)$ becomes a constant.  The ground-state limit is thus recovered since an additional constant on a potential has no physical effect. Though this difference is not close to a constant in the small-$r$ region for larger $\wt$ (Fig. \ref{fig:He:deltavxc:more}), evidence of the step down remains in the shoulder present before the sharp decrease to the ground-state potential. We showed\cite{PYTB14} that the switching point $r\c$ for small values of $\wt$ depends on $\log\wt$, so the $\wt\to0$ limit is achieved slowly as $\wt$ decreases. The large-$\wt$ difference between the ground-state and ensemble XC potentials (Fig. \ref{fig:He:vxcatw}) appears to emerge continuously from the step-like small-$\wt$ behavior, suggesting that the derivative discontinuity is crucial for replication of the bump in $v\xc(\br)$.


With the exact ensemble XC potentials available, we can numerically verify exact conditions of EDFT, such as the virial theorem\cite{N02,Nb95}. With traditionally defined Hartree, its form is similar to its ground-state counterpart\cite{LP85}:
\ben
T_{{\sss C},\wt}[n]=-E\xcw[n]-\int\intd^3r\;n(\br)\br\cdot\nabla v\xcw(\br).
\label{eqn:virialxc}
\een
The virial as defined by Nagy yields the same results as directly calculated kinetic correlation to within 1\%.


\begin{figure}[htbp]
\includegraphics[width=0.9\columnwidth]{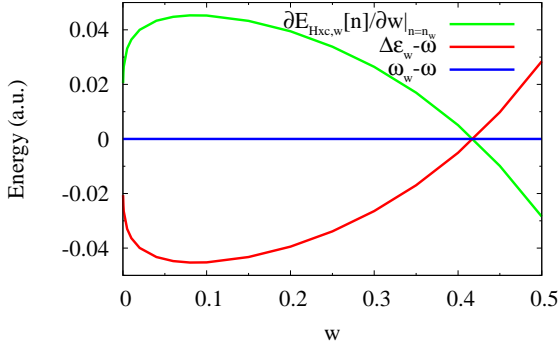}
\caption{Eq. \parref{eqn:theory:exciteng} applied to the exact helium singlet ensemble, demonstrating the exact cancellation of all $\wt$-dependence in KS gaps (red) and corrections to the KS gap (green), leading to no $\wt$ dependence in the calculated optical gap (blue). Gaps are shifted by the true optical gap $\omega$ for ease of comparison. Color online.}
\label{fig:deltaEdeltaeps:exact}
\end{figure}

Eq. \parref{eqn:theory:exciteng} converts the $\wt$-dependent KS transition energies, $\Delta\epsilon_\wt$, into the exact, $\wt$-independent transition frequency. The last term in Eq. \parref{eqn:theory:exciteng} is significant for all values
of $\wt$ and is strongly $\wt$-dependent. Fig. \ref{fig:deltaEdeltaeps:exact} shows the exact cancellation of the $\wt$-dependence as required by Eq. \parref{eqn:theory:exciteng}. If this cancellation is incomplete, as it is in existing approximations, $\wt$-dependent excitation energies will result.

The strong $\wt$-dependence in the exact KS gap $\Delta\epsilon_\wt$ is related to the bumps in the exact XC potentials (Fig. \ref{fig:He:vxcatw}). The bump near $r=2.5$ creates a local confinement effect near the nucleus, shifting the KS eigenvalues upward from the ground-state values. The effect is smaller for the 1s orbital because the 1s orbital density is already small and monotonically decaying at the position of the bump. The KS gap becomes larger as the bump is more prominent, as can be seen in the large-$\wt$ region of Fig. \ref{fig:deltaEdeltaeps:exact}. The sharp change of $\Delta\epsilon_\wt$ in the small-$\wt$ region of Fig. \ref{fig:deltaEdeltaeps:exact} is due to the ensemble derivative discontinuity, since $\Delta v\xc(r)$ effectively creates a bump in the XC potential in the small-$r$ region.

\section{Approximations}
To illustrate the usefulness of these results, we test the few existing approximations to EDFT, including the quasi-local-density approximation (qLDA)\cite{K86,OGKb88}, the single-Slater-determinant ghost-corrected exact exchange (SD)\cite{N98,GPG02}, and the symmetry eigenstate Hartree-exchange (SEHX)\cite{GPG02,PYTB14}. Both SD and SEHX are approximations falling under the overarching work on ghost interactions by Gidopoulos, Papaconstantinou, and Gross\cite{GPG02}, which we denote here as GPG.  The flexibility of GPG lies in its general approach to the description and elimination of ghost interactions introduced by the exchange and traditionally defined Hartree energies.  These ghosts occur when one uses the ensemble density as input into these terms, as there are spurious interactions between the ground and excited states.  If one uses the ensemble definition of Hartree-exchange in Eq. \parref{eqn:EHx}, these ghosts are avoided.  

As a general methodology, GPG can be used in various forms.  When faced with degenerate states, one always has choices about which states to use to describe the system of interest.  Two obvious choices are single- and multi-determinant descriptions.  When the GPG methodology is applied to ensemble Hartree-exchange using symmetry eigenstates with the Krieger-Li-Iafrate approximation\cite{KLI90}, one produces the SEHX approximation. Alternatively, one may choose to use single-determinant states within the GPG methodology.  We show this SD approach alongside the SEHX approximation to clarify the effect of using full eigenstates to describe ensemble ghosts, since previous calculations\cite{N01,TN03,CKZ11,KK13,PP14} can be reevaluated in light of these comparisons.

The general equation of the SEHX energy for an ensemble up to the $I$-th group of degenerate states(`multiplet') is\cite{PYTB14}
\begin{widetext}
\begin{multline}
\label{SEHXeqn}
E\Hx^\text{SEHX}=\int\frac{\intd^3r\intd^3r'}{\abs{\br-\br'}}\Bigg\{\sum_{\mu,\nu>\mu}\big\{n^\text{orb}_\mu(\br)n^\text{orb}_\nu(\br')-\Re[n^\text{orb}_\mu(\br',\br)n^\text{orb}_\nu(\br,\br')]\delta_{\sigma_\mu,\sigma_\nu}\big\}\sum_{i=1}^I\sum_{k=1}^{g_i}\wt_{i,k}\sum_{p=1}^{\tilde{g}_{\tilde{i}}}\abs{C_{i,k,p}}^2f_{\tilde{i},p,\mu}f_{\tilde{i},p,\nu}\\
+\sum_{\substack{\mu,\nu>\mu\\ \kappa,\lambda>\kappa}}[\phi_\mu^*(\br)\phi_\nu^*(\br')\phi_\kappa(\br)\phi_\lambda(\br')\delta_{\sigma_\mu,\sigma_\kappa}\delta_{\sigma_\nu,\sigma_\lambda}-\phi_\mu^*(\br)\phi_\nu^*(\br')\phi_\lambda(\br)\phi_\kappa(\br')\delta_{\sigma_\mu,\sigma_\lambda}\delta_{\sigma_\nu,\sigma_\kappa}]
\sum_{i=1}^I\sum_{k=1}^{g_i}\wt_{i,k}\sum_{p,q\ne p}^{\tilde{g}_{\tilde{i}}}C_{i,k,p}^*C_{i,k,q}\\
\times f_{\tilde{i},p,\mu}f_{\tilde{i},p,\nu}f_{\tilde{i},q,\kappa}f_{\tilde{i},q,\lambda}\prod_{\eta\ne\mu,\nu,\kappa,\lambda}\delta_{f_{\tilde{i},p,\eta},f_{\tilde{i},q,\eta}}\Bigg\},
\end{multline}
\end{widetext}
where $i$ denotes a multiplet; $k$ denotes a specific state in the $i$-th multiplet; $g_i$ is the degeneracy of the $i$-th multiplet; $\tilde{g}_{\tilde{i}}$ is the degeneracy of the corresponding Kohn-Sham (KS) multiplet; $p,q$ denote specific KS single Slater determinants; $\mu,\nu,\kappa,\lambda,\eta$ denote KS orbitals; $\wt_{i,k}$ is the weight of the $k$-th state in the $i$-th multiplet; $C_{i,k,p}$ is the mixing coefficient of the $p$-th determinant to make up the $k$-th state in the $i$-th multiplet; $f_{\tilde{i},p,\mu}$ is the occupation number of the $\mu$-th orbital in the $p$-th determinant of the $\tilde{i}$-th KS multiplet; $\sigma$ denotes spin, $\phi$ denotes KS orbitals; $n^\text{orb}_\mu(\br)$ is the orbital density of the $\mu$-th orbital; and $n^\text{orb}_\mu(\br,\br')=\phi_\mu(\br)\phi_\mu^*(\br')$.  

This form is more explicit than the one given in our previous work\cite{PYTB14}, in order to facilitate use of the SEHX
version of GPG.  Ref.~\cite{GPG02} presents the general framework and a single-determinant example based on the exact exchange OEP formalism of Nagy\cite{N98,Nb98}.  However, the authors use the ensemble Hartree-exchange definition of Eq. \parref{eqn:EHx} and symmetry eigenstates to calculate their reported results.  We have denoted such a procedure as SEHX.  SEHX, as written out here and in Ref. \cite{PYTB14}, yields self-consistent results that agree to within 0.03 eV with those presented in Table I of Ref. \cite{GPG02}, with this difference assumed to be due to numerical differences in implementation.


\section{Approximate Results}

\begin{figure}[htbp]
\includegraphics[width=0.9\columnwidth]{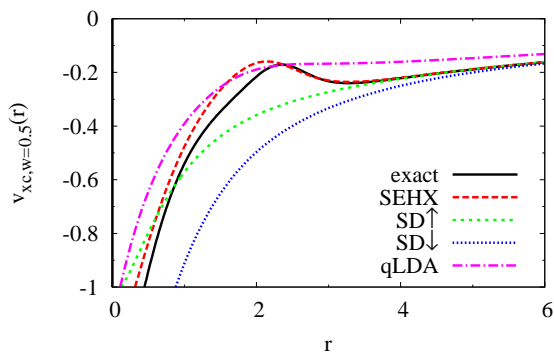}
\caption{The exact and approximated $v\xc(\br)$ for the helium singlet equiensemble. The approximated $v\xc$'s are evaluated using the exact ensemble density as input.}
\label{fig:He:approx}
\end{figure}

Comparison of exact and approximate quantities exposes differences in single- and multi-determinant approximations, as well as the shortcomings both share.  Fig. \ref{fig:He:approx} shows exact and approximate XC potentials using the exact ensemble density. Both the SD and the SEHX are OEPs, which guarantees their correct $-1/r$ asymptotic behavior in the XC potential (Fig. \ref{fig:He:approx}). However, only the SEHX potential shows the large $\wt$ bump and recovers the general shape of the exact $v_{{\sss XC},\wt}(r)$.

\begin{figure}[htbp]
\includegraphics[width=0.9\columnwidth]{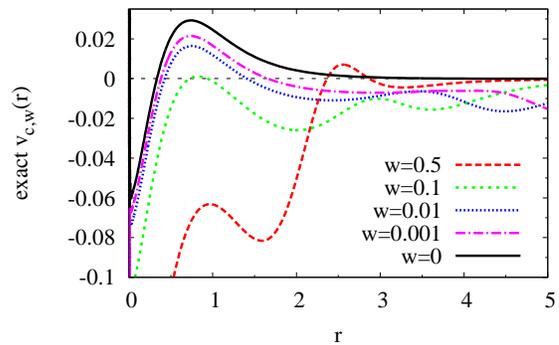}
\caption{The exact $v\c(r)$ for the helium singlet equiensemble shows two upward bumps and does not depend on the definition of the Hartree potential used. These are obtained by subtracting SEHX $v_{{\sss X},\wt}(r)$ of the exact ensemble density from the exact $v_{{\sss XC},\wt}(r)$.}
\label{fig:He:correct:vc}
\end{figure}

The correlation potential $v_{{\sss C},\wt}(r)$ displays two distinct bumps, shown in Fig. \ref{fig:He:correct:vc}. The $w=0$ correlation potential matches perfectly with the exact ground-state correlation potential in Ref. \cite{UG94}. The first bump at about $r=1$ also exists in the ground-state $v\c(r)$, while the second bump at about $r=2.5$, which vanishes rapidly as $\wt$ decreases, is unique to EDFT.

\begin{figure}[htbp]
\includegraphics[width=0.9\columnwidth]{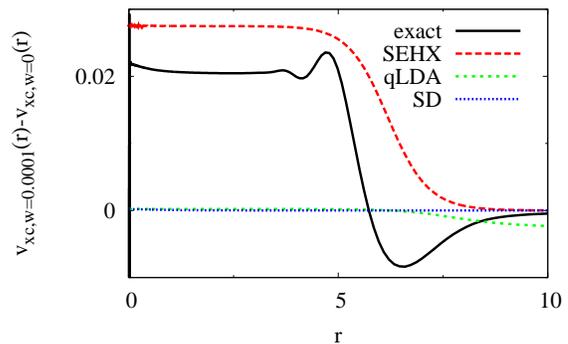}
\caption{Self-consistent $\Delta v\xc(r)$ of various approximations at $\wt=0.0001$. Only SEHX (dotted red) replicates a shift similar to that of the exact curve (solid black). Color online.}
\label{fig:He:deltavxc:0001}
\end{figure}

Fig. \ref{fig:He:deltavxc:0001} shows that, in the small $\wt$ region, only SEHX generates a step-like form for the ensemble derivative discontinuity.  The SEHX XC potential is also the only approximation that has the characteristic bump of the exact XC potential. Both SEHX and SD are OEP methods, but the former satisfies the exact condition of the ensemble derivative discontinuity, while the latter does not.  The SEHX potential is obtained by applying the KLI approximation\cite{KLI90} to the optimized effective potential (OEP) equation\cite{N98}. Equations for $v_{{\sss HX},\wt}^\text{SEHX}(\br)$ of the helium singlet bi-ensemble are given in Eqs. 41 - 43 of Ref. \cite{PYTB14}.   

To understand the absence of the derivative discontinuity in SD, we compare the small-$\wt$ behavior of both SD and SEHX\cite{PYTB14}.  The SD potential for the spin-up electron is
\ben
\begin{split}
v_{{\sss HX\uparrow},\wt}^\text{SD}(\br)=&\Big\{(1-\wt)n_{1\uparrow}^\text{orb}(\br)[v_{1\uparrow}(\br)+\bar{v}_{{\sss HX1\uparrow},\wt}-\bar{v}_{1\uparrow}]\\
&+\wt n_{2\uparrow}^\text{orb}(\br)[v_{2\uparrow}(\br)+\bar{v}_{{\sss HX2\uparrow},\wt}-\bar{v}_{2\uparrow}]\Big\}/n_{\uparrow,\wt}(\br),
\end{split}
\een
where $n_{\uparrow,\wt}(\br)=(1-\wt)n_{1\uparrow}^\text{orb}(\br)+\wt n_{2\uparrow}^\text{orb}(\br)$, and
\ben
v_{1\uparrow}(\br)=v_{2\uparrow}(\br)=\int\frac{\intd^3r'}{\abs{\br-\br'}}n^\text{orb}_{1\downarrow}(\br').
\een
Barred quantities are defined
\ben
\bar{v}_j=\int\intd^3r\;v_j(\br)n_j^\text{orb}(\br),
\een
so that $\bar{v}_{{\sss HX1\uparrow},\wt}$, for instance, is the expectation value of the spin-up HX potential with respect to $n_{1\uparrow}^\text{orb}(\br)$.

Comparing the SEHX\cite{PYTB14} and SD expressions for the HX potentials makes the disappearance of the derivative discontinuity in the SD approximation clear.  When $\wt$ is very small, in the region where $r$ is smaller than a certain $r\c$, $n_\wt(\br)$ is dominated by the $(2-\wt)n_1^\text{orb}(\br)$ term (see Eq. 41 of Ref. \cite{PYTB14}). In the $r>r\c$ region, however, it is dominated by the $w n_2^\text{orb}(\br)$ term due to the slower decay of $n_2^\text{orb}(\br)$. Thus, when $\wt$ is very small, we have
\ben
v_{{\sss HX},\wt\approx 0}^\text{SEHX}(\br)\approx\left\{
\begin{array}{ll}
v_1(\br)+\bar{v}_{\sss HX1}-\bar{v}_1, & r<r\c,\\
v_2(\br)+\bar{v}_{\sss HX2}-\bar{v}_2, & r>r\c,
\end{array}
\right.
\label{eqn:SEHX:v1v2}
\een
and
\ben
v_{{\sss HX\uparrow},\wt\approx 0}^\text{SD}(\br)\approx\left\{
\begin{array}{ll}
v_{1\uparrow}(\br)+\bar{v}_{{\sss HX1\uparrow},\wt}-\bar{v}_{1\uparrow}, & r<r\c,\\
v_{2\uparrow}(\br)+\bar{v}_{{\sss HX2\uparrow},\wt}-\bar{v}_{2\uparrow}, & r>r\c,
\end{array}
\right.
\label{eqn:sdGPG:v1v2}
\een

For any $\wt$, $v_{1\uparrow}(\br)=v_{2\uparrow}(\br)$, so the SD approximation yields the same behavior at large or small $\wt$. In contrast, when $\wt$ is very small within the SEHX approximation,
\ben
v_1(\br)\approx\int\frac{\intd^3r'}{\abs{\br-\br'}}n_1(\br'),
\een
and
\ben
\label{fdefn}
\begin{split}
v_2(\br)&=\int\frac{\intd^3r'}{\abs{\br-\br'}}\left[n_1^\text{orb}(\br')+\frac{\phi_1^*(\br)\phi_2^*(\br')\phi_1(\br')}{\phi_2^*(\br)}\right]\\
&=v_1(\br)+f(\br).
\end{split}
\een
$v_1(\br)$ and $v_2(\br)$ therefore have a finite difference even at $\wt=0$. We have shown that $r\c\approx -0.621\ln\wt$ in Ref. \cite{PYTB14}, so the constant terms in Eq. \parref{eqn:SEHX:v1v2} are
\ben
\begin{split}
\bar{v}_{\sss HX1}(\br)-\bar{v}_1(\br)&=\int\intd^3r\;n_1^\text{orb}(\br)[v_{{\sss HX},\wt\approx 0}^\text{SEHX}(\br)-v_1(\br)]\\
&\approx\int\intd\Omega\int_{r\c}^\infty\intd r\;n_1^\text{orb}(\br)f(\br),
\end{split}
\label{eqn:SEHX:barv1}
\een
because the integrand vanishes when $r<r\c$ and $\wt$ is small.  Similarly,
\ben
\bar{v}_{\sss HX2}(\br)-\bar{v}_2(\br)\approx-\int\intd \Omega \int_0^{r\c}\intd r\;n_2^\text{orb}(\br)f(\br).
\label{eqn:SEHX:barv2}
\een
Eq. \ref{fdefn} shows that $f(\br)$ decreases rapidly as $\br$ increases, since $\phi_1(\br)$ decays faster asymptotically than $\phi_2(\br)$. Since $f(\br)$ is a part of $v_2(\br)$, which only dominates the large-$r$ behavior of $v_{{\sss HX},\wt\approx 0}^\text{SEHX}(\br)$, the difference between the large-$r$ and small-$r$ behaviors of $v_{{\sss HX},\wt\approx 0}^\text{SEHX}(\br)$ is due to the constant terms in Eqs. \parref{eqn:SEHX:barv1} and \parref{eqn:SEHX:barv2}. In the $w\to0$ limit, Eq. \parref{eqn:SEHX:barv1} vanishes, and Eq. \parref{eqn:SEHX:barv2} approaches a finite negative value. The additive constant in the HX potential obtained needs to be determined by matching with the known $1/r$ behavior, and the resulting potential would show the upward ensemble derivative discontinuity step illustrated in Fig. \ref{fig:He:deltavxc:exact}. Since both $\bar{v}_{{\sss HX1\uparrow},\wt}-\bar{v}_{1\uparrow}$ and $\bar{v}_{{\sss HX2\uparrow},\wt}-\bar{v}_{2\uparrow}$ vanish in the $\wt\to0$ limit, there is no ensemble derivative discontinuity for SD.

\begin{figure}[htbp]
\includegraphics[width=0.9\columnwidth]{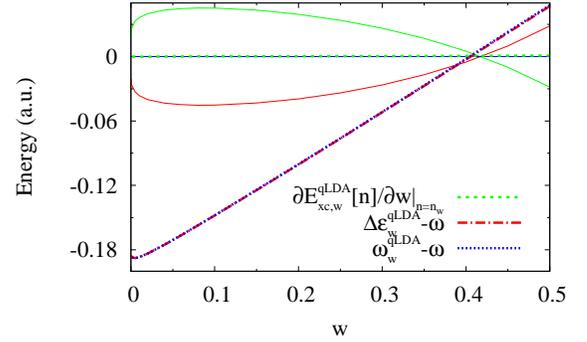}
\caption{Eq. \parref{eqn:theory:exciteng} applied to self-consistent quasi-LDA results. The correction to the quasi-LDA KS gap (dashed green) is not 0, but it is too small to be noticed on this scale.  This correction is inadequate to cancel the $\wt$-dependence in the qLDA KS gap (dashed red), resulting in inaccurate, $\wt$-dependent calculated optical gaps (dashed blue). The gaps have been shifted in this figure by the optical gap $\omega$ for easier comparison, and the exact results of Fig. \ref{fig:deltaEdeltaeps:exact} are also shown for context.  Color online.}
\label{fig:deltaEdeltaeps:qLDA}
\end{figure}

\begin{figure}[htbp]
\includegraphics[width=0.9\columnwidth]{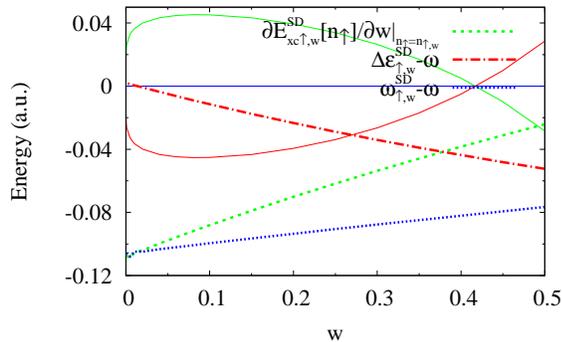}
\caption{Eq. \parref{eqn:theory:exciteng} applied to self-consistent SD results. The spin-up SD KS gap (dashed red) is insufficiently corrected by the SD corrections to the KS gap (dashed green), yielding calculated optical gaps that are too small (dashed blue). Though the $\wt$-dependence is less severe than for qLDA, it is still non-negligible. The gaps have been shifted in this figure by the optical gap $\omega$ for easier comparison, and the exact results of Fig. \ref{fig:deltaEdeltaeps:exact} are also shown for context. Color online.}
\label{fig:deltaEdeltaeps:SD}
\end{figure}

\begin{figure}[htbp]
\includegraphics[width=0.9\columnwidth]{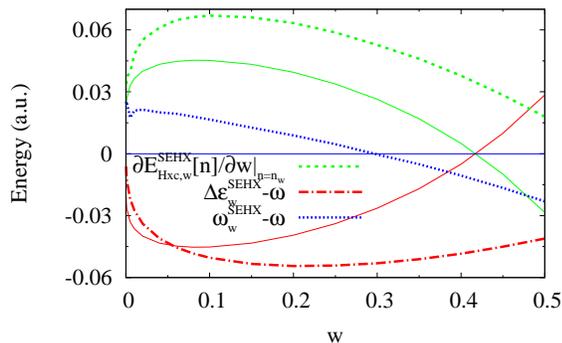}
\caption{Eq. \parref{eqn:theory:exciteng} applied to approximate self-consistent SEHX results. SEHX produces far less variation in calculated excitation energies with $\wt$ (dashed blue), which appears to be the result of its ensemble derivative discontinuity. This produces approximate KS gaps (dashed red) and KS gap corrections (dashed green) that most closely resemble the exact curves in overall shape. The exact results (as in Fig. \ref{fig:deltaEdeltaeps:exact}) are also shown for context. The gaps have been shifted in this figure by the optical gap $\omega$ for easier comparison, and the exact results of Fig. \ref{fig:deltaEdeltaeps:exact} are also shown for context. Color online.}
\label{fig:deltaEdeltaeps:SEHX}
\end{figure}

Figs. \ref{fig:deltaEdeltaeps:qLDA}, \ref{fig:deltaEdeltaeps:SD}, and \ref{fig:deltaEdeltaeps:SEHX} demonstrate that qLDA, SD, and SEHX approximations are unable to generate $\wt$-independent excitation energies. The less severe $\wt$-dependence of the SEHX KS gap is due to its closer replication of the exact ensemble derivative discontinuity, though the SEHX cancellation of excitation energy $\wt$-dependence is not exact. Fig. \ref{fig:He:approx} shows that the position of the large $\wt$ bump of SEHX is at smaller $r$ values than the exact one. This agrees with the less rapid change of the SEHX KS gap in the large-$\wt$ region. In Fig. \ref{fig:deltaEdeltaeps:SEHX}, the sharp change of the SEHX KS gap in the small-$\wt$ region is similar to that of the exact ensemble, which is due to the bump created by the step in $\Delta v\xc$. qLDA and SD potentials have neither the large-$\wt$ bump nor the small-$\wt$ derivative discontinuity step, so the $\wt$-dependencies of their KS gaps are very different from the exact one. Comparing to Figs. \ref{fig:He:vxcatw} and \ref{fig:He:approx}, the $r=2.5$ bump in the correlation potential (Fig. \ref{fig:He:correct:vc}) fixes the position of the bump in the exchange-only (SEHX) potential, and thereby sets the $\wt$-dependence of the KS gap and its correction.

\section{Conclusion}

This work provides a method for inverting ensemble densities, so that the resulting exact ensemble KS systems can be used as references for developing approximated EDFT functionals. We show the density-inversion method for spherically-symmetric systems in this paper, but it is not difficult to generalize the method for other types of systems. We have tested the density-inversion method in cylindrically-symmetric systems and it also yields good results\cite{PYTB14}. For systems with lower symmetry, the real-space approach shown in this paper would not yield accurate results without a massive grid point set. Though expression in a basis set may solve this problem, further study is required to determine the effect this would have on the density-inversion method's stability and performance.

We applied the density-inversion method on the helium singlet bi-ensemble for its simplicity.  This exposes the continuous emergence of the exact XC potential bump from the ensemble derivative discontinuity and facilitates comparison with approximations. The singlet bi-ensemble is by no means the limit of the applicability of the density-inversion method, however. In Ref. \cite{PYTB14}, we apply the method to ensembles of various real and model 2-electron systems, in which it retains the numerical stability and accuracy seen in this paper. This work illustrates that EDFT properties deviate from ground-state DFT ones in previously unseen ways. Also, some exact conditions, such as Eq. \parref{eqn:theory:exciteng}, do not suggest obvious methods for their satisfaction by approximations. Of the approximations we tested, the SEHX version of GPG, the only one with an ensemble derivative discontinuity, generated the most accurate XC potentials and excitation energies. These complications make developing a good EDFT functional considerably harder than in ground state, and we hope the exact results shown in this work can alleviate some burden on EDFT developers.

We thank Nikitas Gidopoulos for providing very helpful clarifications regarding Ref. \cite{GPG02}. Z.-H.Y. thanks Yu Zhang and Daniel Jensen for very helpful discussions on density inversion problems. Z.-H.Y. and C.A.U. are supported by NSF grant No. DMR-1005651. A.P.J. is supported by DOE grant DE-FG02-97ER25308. J.R.T. and R.J.N. acknowledge financial support from the Engineering and Physical Sciences Research Council (EPSRC) of the UK. K.B. supported by DOE grant DE-FG02-08ER46496.


\begin{thebibliography}{10}

\bibitem{B12}
K.~Burke,
\newblock J. Chem. Phys., {\bf 136}, 150901 (2012).

\bibitem{PBE96}
J.~P. Perdew, K.~Burke, and M.~Ernzerhof,
\newblock Phys. Rev. Lett., {\bf 77}, 3865 (1996).
\newblock {\it ibid.} {\bf 78}, 1396(E) (1997).

\bibitem{B93}
A.~D. Becke,
\newblock J. Chem. Phys., {\bf 98}, 5648 (1993).

\bibitem{UG94}
C.~J. Umrigar and X.~Gonze,
\newblock Phys. Rev. A, {\bf 50}, 3827 (1994).

\bibitem{LB94}
R.~van Leeuwen and E.~J. Baerends,
\newblock Phys. Rev. A, {\bf 49}, 2421 (1994).

\bibitem{GLB95}
O.~V. Gritsenko, R.~van Leeuwen, and E.~J. Baerends,
\newblock Phys. Rev. A, {\bf 52}, 1870 (1995).

\bibitem{PVW03}
K.~Peirs, D.~Van~Neck, and M.~Waroquier,
\newblock Phys. Rev. A, {\bf 67}, 012505 (2003).

\bibitem{MMNG12}
M.~A.~L. Marques, N.~T. Maitra, F.~M.~S. Nogueira, E.~K.~U. Gross, and
  A.~Rubio, eds.,
\newblock {\em {F}undamentals of {T}ime-{D}ependent {D}ensity {F}unctional  {T}heory},
\newblock {L}ecture {N}otes in {P}hysics (Springer, Berlin, 2012).

\bibitem{U12}
C.~A. Ullrich,
\newblock {\em {T}ime-{D}ependent {D}ensity-{F}unctional {T}heory: {C}oncepts
  and {A}pplications},
\newblock (Oxford University Press, Oxford, 2012).

\bibitem{JPCA10}
D.~Jacquemin, E.~A. Perp\`ete, I.~Ciofini, C.~Adamo, R.~Valero, Y.~Zhao, and
  D.~G. Truhlar,
\newblock J. Chem. Theory Comput., {\bf 6}，2071 (2010).

\bibitem{AGB03}
H.~Appel, E.~K.~U. Gross, and K.~Burke,
\newblock Phys. Rev. Lett., {\bf 90}, 043005 (2003).

\bibitem{Gb99}
A.~G\"orling,
\newblock Phys. Rev. A, {\bf 59}, 3359 (1999).

\bibitem{LN99}
M.~Levy and {\'A}.~Nagy,
\newblock Phys. Rev. Lett., {\bf 83}, 4361 (1999).

\bibitem{T79}
A.K. Theophilou,
\newblock J. Phys. C, {\bf 12}, 5419 (1979).

\bibitem{HT85}
N.~Hadjisavvas and A.~K. Theophilou,
\newblock Phys. Rev. A, {\bf 32}, 720 (1985).

\bibitem{T87}
A.~K. Theophilou,
\newblock in {\em The Single-Particle
  Density in Physics and Chemistry}, edited by N.~H. March and B.~M. Deb (Academic press, London, 1987).

\bibitem{GOKb88}
E.~K.~U. Gross, L.~N. Oliveira, and W.~Kohn,
\newblock Phys. Rev. A, {\bf 37}, 2805 (1988).

\bibitem{GOK88}
E.~K.~U. Gross, L.~N. Oliveira, and W.~Kohn,
\newblock Phys. Rev. A, {\bf 37}, 2809 (1988).

\bibitem{OGKb88}
L.~N. Oliveira, E.~K.~U. Gross, and W.~Kohn,
\newblock Phys. Rev. A, {\bf 37}, 2821 (1988).

\bibitem{N95}
{\'A}.~Nagy,
\newblock Int. J. Quantum Chem., {\bf 56}(S29), 297 (1995).

\bibitem{N98}
{\'A}~Nagy,
\newblock Int. J. Quant. Chem., {\bf 69}, 247 (1998).

\bibitem{N01}
{\'A}~Nagy,
\newblock J. Phys. B: At. Mol. Opt. Phys., {\bf 34}, 2363 (2001).

\bibitem{GPG02}
N.~I. Gidopoulos, P.~G. Papaconstantinou, and E.~K.~U. Gross,
\newblock Phys. Rev. Lett., {\bf 88}, 033003 (2002).

\bibitem{PGP13}
E.~Pastorczak, N.~I. Gidopoulos, and K.~Pernal,
\newblock Phys. Rev. A, {\bf 87}, 062501 (2013).

\bibitem{PP14}
E.~Pastorczak and K.~Pernal,
\newblock J. Chem. Phys., {\bf 140}, 18A514 (2014).

\bibitem{JG89}
R.O. Jones and O.~Gunnarsson,
\newblock Rev. Mod. Phys., {\bf 61}, 689 (1989).

\bibitem{drake_94}
G.~W.~F. Drake and Z.-C. Yan,
\newblock Chem. Phys. Lett., {\bf 229}, 486 (1994).

\bibitem{nistor_04}
R.~A. Nistor,
\newblock M.Sc. Thesis, University of Windsor, Canada (2004).

\bibitem{L95}
M~Levy,
\newblock Phys. Rev. A, {\bf 52}, R4313 (1995).

\bibitem{note1}
There appears to be a sign error in Eq. (16) of Ref. \cite{L95}: the two terms on the right-hand side should be swapped.

\bibitem{PYTB14}
A.~Pribram-Jones, Z.-H. Yang, J.~R. Trail, K.~Burke, R.~J. Needs, and C.~A. Ullrich,
\newblock J. Chem. Phys., {\bf 140}, 18A541 (2014).

\bibitem{N02}
{\'A}~Nagy,
\newblock Acta Phys. Chim. Debrecina, {\bf 34-35}, 99 (2002).

\bibitem{Nb95}
{\'A}.~Nagy,
\newblock {\em International Journal of Quantum Chemistry}, 56(4):225--228,
  1995.

\bibitem{LP85}
M.~Levy and J.~P. Perdew,
\newblock Phys. Rev. A, {\bf 32}, 2010 (1985).

\bibitem{K86}
W.~Kohn,
\newblock Phys. Rev. A, {\bf 34}, 737 (1986).

\bibitem{KLI90}
J.~B. Krieger, Y.~Li, and G.~J. Iafrate,
\newblock Phys. Lett. A, {\bf 146}, 256 (1990).

\bibitem{TN03}
F~Tasn{\'a}di and {\'A}~Nagy,
\newblock J. Phys. B: At. Mol. Opt. Phys., {\bf 36}, 4073 (2003).

\bibitem{CKZ11}
John Cullen, Mykhaylo Krykunov, and Tom Ziegler,
\newblock Chem. Phys., {\bf 391}, 11 (2011).

\bibitem{KK13}
Eli Kraisler and Leeor Kronik,
\newblock Phys. Rev. Lett., {\bf 110}, 126403 (2013).

\bibitem{Nb98}
{\'A}.~Nagy,
\newblock Int. J. Quant. Chem., {\bf 70}, 681 (1998).

\end{thebibliography}
\end{document}